\documentclass{elsart}

\usepackage{graphicx}
\begin{document}
\title{On Axial and Plane--Mirror Inhomogeneities in the WMAP3 Cosmic
Microwave Background Maps}
\author{V.G.Gurzadyan$^{1,2}$, A.A.Starobinsky$^3$, A.L.Kashin$^1$,
H.Khachatryan$^{1,2}$, G.Yegorian$^{1,2}$}
\address{(1) Yerevan Physics Institute, Yerevan, Armenia\\
(2) ICRANet, ICRA,  University ``La Sapienza'', Rome, Italy\\
(3) Landau Institute for Theoretical Physics, Moscow, 119334,
Russia}

{\bf Abstract.} We study inhomogeneities in the distribution of the
excursion sets in the Cosmic Microwave Background (CMB) temperature
maps obtained by the three years survey of the Wilkinson Microwave
Anisotropy Probe (WMAP). At temperature thresholds $|T|<90 \mu K$,
the distributions of the excursion sets with over 200 pixels are
concentrated in two regions, nearly at the antipodes, with galactic
coordinates $l= 94^\circ.7,\, b= 34^\circ.4$ and $l= 279^\circ.8,\,
b= -29^\circ.2$. The centers of these two regions drift towards the
equator when the temperature threshold is increased. The centers are
located close to one of the vectors of $\ell =3$ multipole. The two
patterns of the substructures in the distribution of the excursion
sets are mirrored, with $\chi^2=0.7-1.5$. There is no obvious origin
of this effect in the noise structure of WMAP, and there is no
evidence for a dependence on the galactic cut. Would this effect be
cosmological, it could be an indication of an anomalously large
component of horizon-size density perturbations, independent of one
of the spatial coordinates, and/or of a non-trivial slab-like
spatial topology of the Universe.

\section{Introduction}

The power spectrum of CMB has provided essential information on the
cosmological parameters \cite{DB1,Sp}. Another important source of
information are CMB temperature maps which are particularly useful
for the study of possible non-Gaussianity signals of various nature.
Among the latter are the alignments of low multipoles, claimed as
anomalous with respect to Gaussian distribution (e.g.
\cite{Oliv,Copi1,Copi2,Erik,Copi3,Dennis,Hell}). In the present
paper we analyse axial inhomogeneities in the distribution of the
excursion sets in the WMAP3 maps; these can be related to the
alignments mentioned above. We have estimated the centers of
excursion sets at given temperature thresholds and for a given pixel
count. The distribution of the centers of the excursion sets
revealed inhomogeneities, with mirroring features, which we analyse
below. Though the Galactic and interplanetary contamination can be
the major contributor here, and hence the non-cosmological nature of
the low multipoles, we nevertheless discuss briefly the principal
conditions for arising of the mirroring from topological properties
of the Universe.

\section{Excursion sets}

For this analysis we used the 94 GHz (3.2mm) WMAP 3-year maps. These
maps feature the highest angular resolution, and are less polluted
than lower frequency channels by synchrotron radiation of our
Galaxy. The $|b|<20^o$ region has been excluded as usual, to
minimize effects due to the Galactic disk.

The algorithm for defining of the center of the pixelized excursion
sets, and of their other parameters, has been described in
\cite{Gur1,Gur2}; temperature independent ellipticity in the
excursion sets was found in Boomerang's and WMAP's maps, and had
been found in the COBE maps as well \cite{GT}. The algorithm is
based on Cartan's proof  that maximally compact subgroups of Lie
groups are always conjugate \cite{Kar}, i.e. due to the conditions
of existence of the center of mass of the given set of points. For
Riemannian manifolds the center of mass of the points $\{x_i\},
i=1,...,n $ is defined as
\begin{equation}
x=\frac{1}{n}(x_1+x_2+...+x_n),
\end{equation}
or that the mass center is the point where the following function is minimized:
\begin{equation}
y \rightarrow \sum_{1}^{n}d(x,x_i)^2.
\end{equation}
Though for present purposes the temperature weighting is not important,
this procedure in principle enables such generalization assigning a
weight $T_i$ to each pixel of coordinate $x_i$.

Without loosing generality consider a compact subset $A\in M$ with a mass distribution $da$
on $A$ with normalized mass 1. Cartan's theorem states the existence of the mass center if the
function
\begin{equation}
f:m \in M \rightarrow \frac{1}{2} \int d(m,a)^2da
\end{equation}
is convex, achieves a unique minimum at the mass center of $A$ for the mass
distribution $da$. The mass center, then, is the unique point of vanishing of
the gradient vector field i.e. of the linear connection or of the covariant
derivative $\nabla f(x)$.

We used this algorithm to obtain the centers of excursion sets of
various pixel count and temperature intervals.

\section{The Inhomogeneities}

Our analysis has shown that the distribution of the centers of the excursion sets,
for pixel counts larger than 100 and 200, reveals anomalous properties. Namely,
starting from the temperature thresholds about $|T| = 80 \mu K$ the centers of the
excursion sets are located in the Northern and Southern hemispheres, and denoted
as "A" and "B", respectively; their coordinates for pixel counts larger than
200-pixel at $|T|=90 \mu K$ are
$$
l= 94^\circ.7,\,\, b= 34^\circ.4\,\, (A);
l= 279^\circ.8,\,\, b= -29^\circ.2\,\, (B),
$$
so that they are nearly antipodal.

One can analyze compare the location of A and B to the multipoles of CMB anisotropy
using the Maxwellian vectors of the multipoles \cite{Copi1,Copi2,Copi3}. While the
coefficients $a_{\ell m}$ are describing the representation
of the temperature by spherical harmonics
\begin{equation}
\frac{\Delta T_{\ell }(\hat{n})}{T}=\sum_{m=-\ell }^{\ell }a_{\ell m}Y_{\ell
m}(\hat{n})
\end{equation}%
with $<a_{\ell m}>=0$
and for Gaussian functions are defining the angular power spectrum
\begin{equation}
C_{\ell }\equiv \frac{1}{2\ell +1}\sum_{m=-\ell }^{\ell }\left\vert a_{\ell
m}\right\vert ^{2},
\end{equation}%
the Maxwellian multipole unit vectors $\{\hat{v}^{(\ell ,i)}\mid
i=1,...,\ell \}$ and a scalar coefficient $A^{(\ell )}$ characterize
the series expansion of a function on a sphere.
The algorithm in \cite{Hell,Dennis} was used to compute the multipole vectors reducing
the problem to finding of the roots of a polynomial.

Figure 1 shows the position of the A and B with respect the vectors
of multipoles $\ell=1-4$. It is seen that A and B define a direction
close to one of the vectors of $\ell =3$ multipole, drifting towards
the equator with the increase of the temperature.

We also calculated the sum vector of multipoles $\ell=2-8$
as shown in Figure 2 (here the module of each vector was weighted by
$1/\ell (\ell +1)$): the open star marks the sum of $\ell =2-8$ multipole
vectors, the black star denotes those of $\ell =1-8$ i.e. together with the
contribution of the dipole. We see that centers A,B are not near either of them.

To probe the dependence of the inhomogeneities on low multipoles, we
removed such components from the WMAP 3 year maps using the "anafast" and "synfast"
functions from Healpix package \cite{Healpix}.

The increase of the number of pixels in the excursion sets with the
radius from A and B is shown in Figure 3, upper plot. The middle and
lower plots in Figure 3 exhibit the same but with extracted
multipoles $\ell =2$ and $\ell =2,3$, respectively.

\section{Mirroring}

Figures 4-6 show the distances of A and B from the dipole, from the closest pole of multipole $\ell =3$ and from the sum of the poles of $\ell =2-8$ multipoles, all as a function of the temperature threshold.
The middle and lower panels of Figure 4 show the dipole-(A,B) distances from multipoles $\ell =2$ and $\ell =2,3$ respectively.

The locations of A and B are not close to those of the cold spot \cite{Vielva}.

The role of scan inhomogeneities and noise is tested by analyzing in the same way the difference of the maps from independent radiometers (A-B). For the studied excursion sets, in the noise map (A-B)a signal-to-noise ratio around 4:1 is found; the excursion sets in this case do not show any of the properties described above for the 94 GHz (A+B sum) map.

\begin{figure}
\begin{center}
\includegraphics[height=8cm,width=12cm]{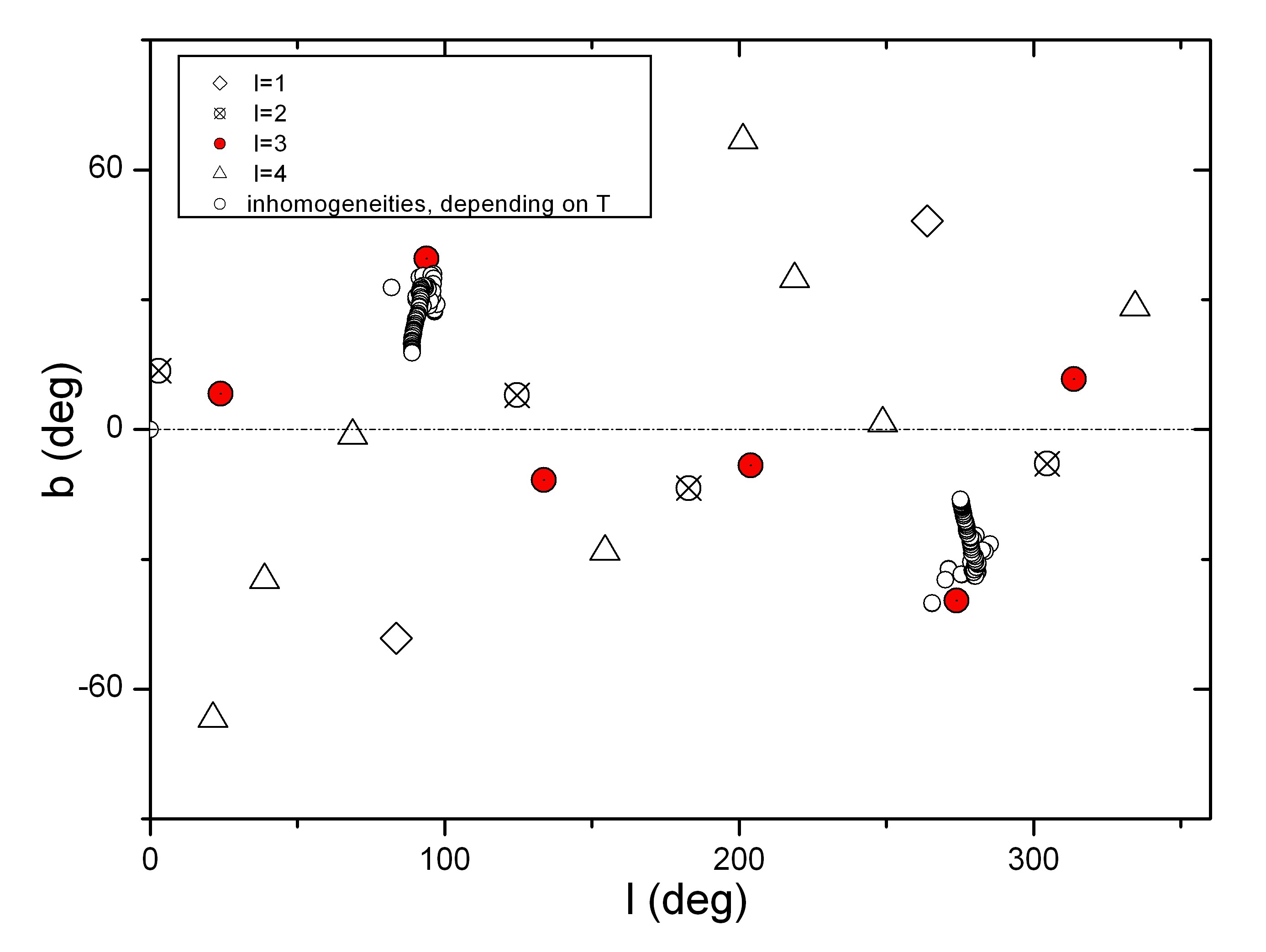}
\caption{Drift of the centers A and B with the increase of the temperature threshold interval and the multipole vectors $\ell = 1-4$.}
\end{center}
\end{figure}

\begin{figure}
\begin{center}
\includegraphics[height=10cm,width=14cm]{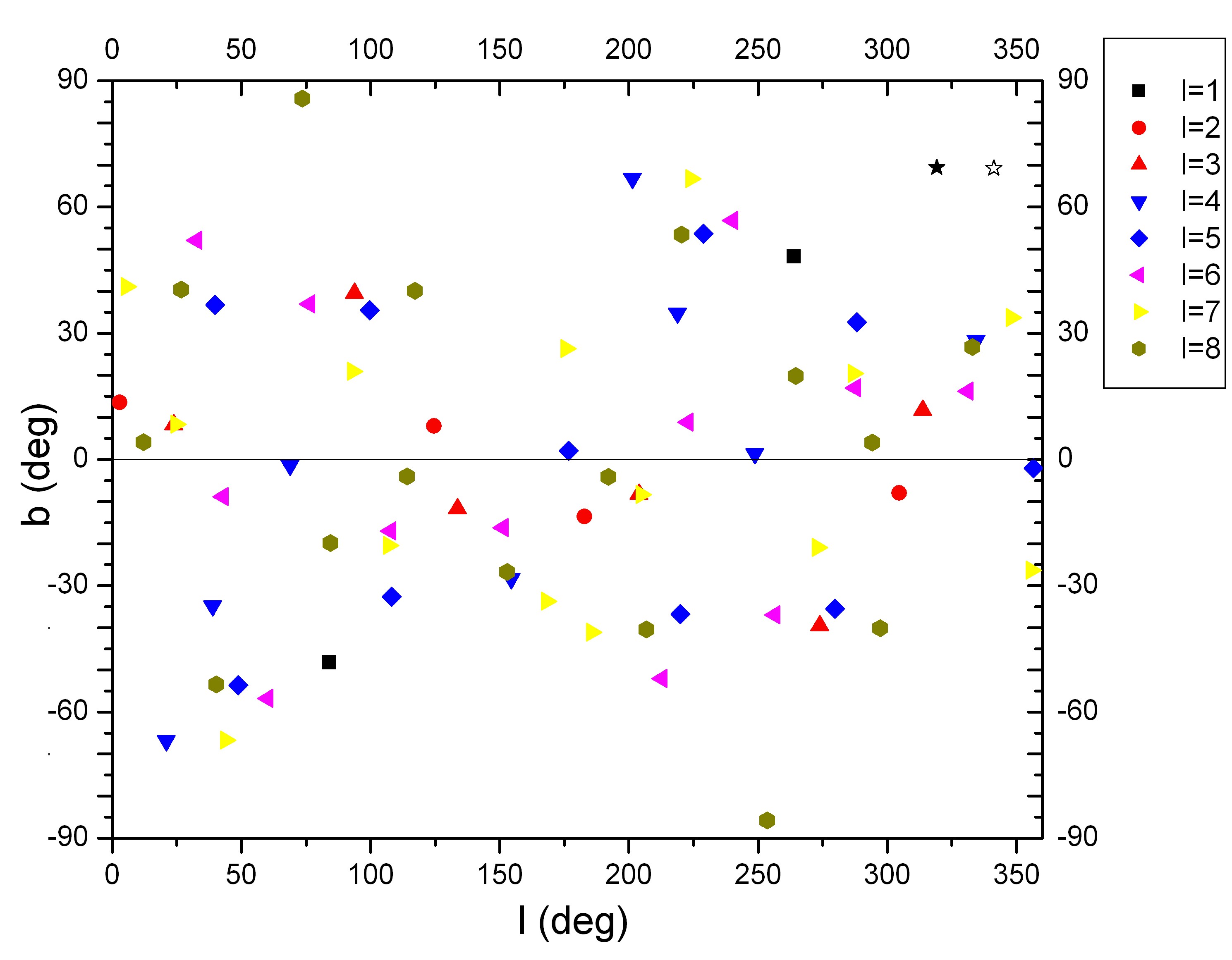}
\caption{Multipole vectors, $\ell =1-8$, and the sum vectors, $\ell =1-8$ (black star) and $\ell =2-8$ (open star).}
\end{center}
\end{figure}

\begin{figure}
\begin{center}
\includegraphics[height=10cm,width=8cm]{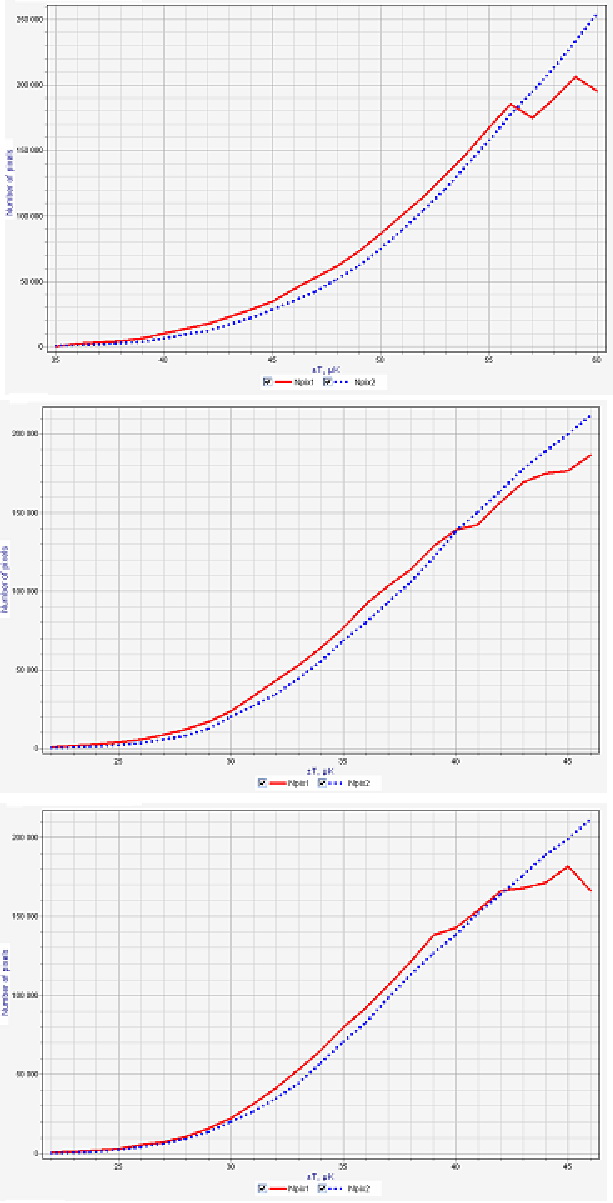}
\caption{Pixel numbers of the excursion sets vs the radius from A and B (dotted line) at temperature step
1 $\mu K$ (upper plot). The middle and lower plots are with $\ell =2$ and $\ell =2,3$ multipoles removed, respectively.}
\end{center}
\end{figure}

\begin{figure}
\begin{center}
\includegraphics[height=10cm,width=8cm]{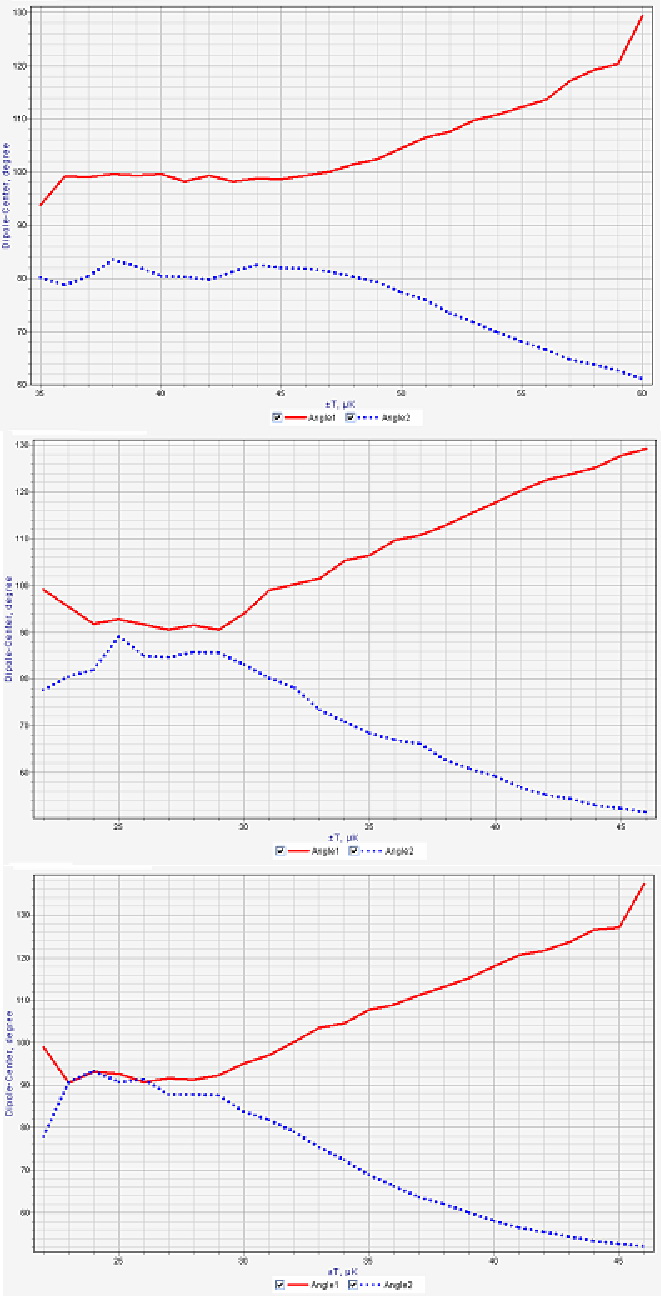}
\caption{The distance of the dipole apexes from A,B; the rest as is in previous figure.
The $\chi^2$ for mirroring is 1.50, 1.48, 4.81, respectively.}
\end{center}
\end{figure}

\begin{figure}
\begin{center}
\includegraphics[height=5cm,width=8cm]{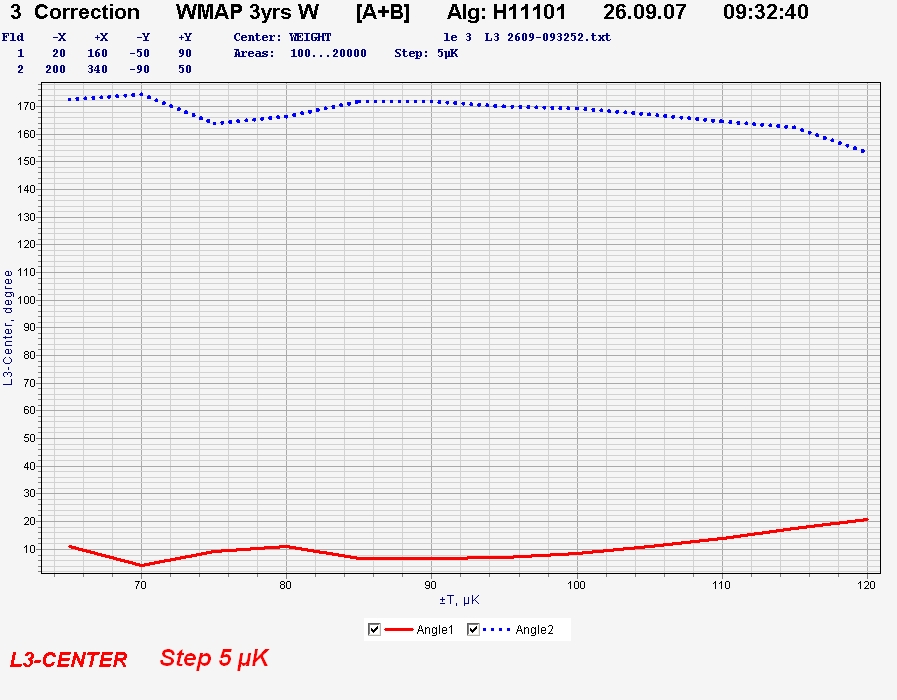}
\caption{The distance of $\ell =3$ (closest vector) from A,B; mirorring $\chi^2=0.74$.}
\end{center}
\end{figure}

\begin{figure}
\begin{center}
\includegraphics[height=5cm,width=8cm]{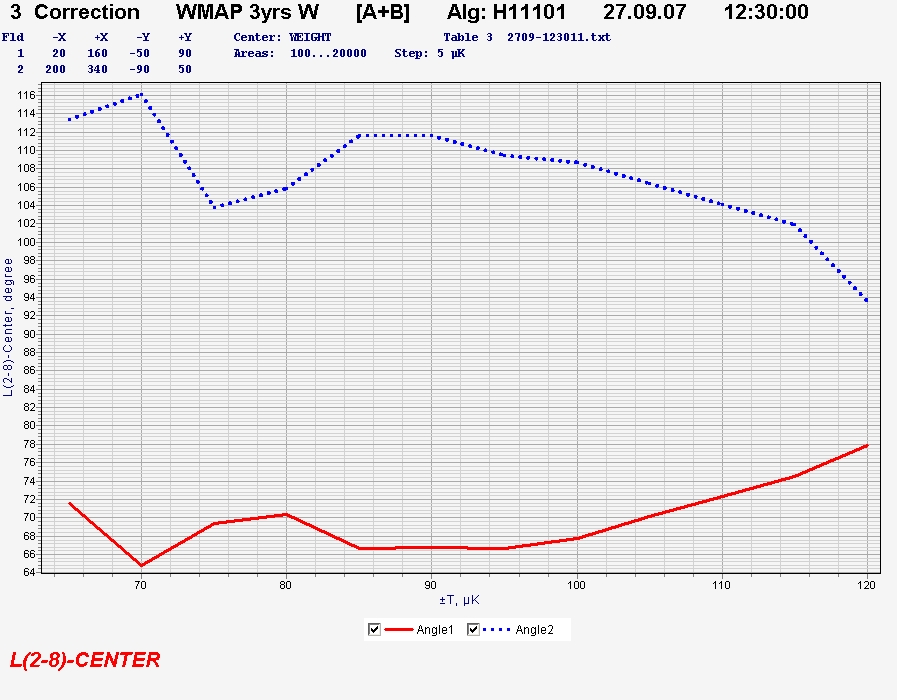}
\caption{The distance of $\ell =2-8$ sum vector from A,B; mirroring $\chi^2=0.78$.}
\end{center}
\end{figure}

\section{Discussion}

The centers of excursion sets have nearly antipodal locations. The inhomogeneity
centers are close to one of $\ell =3$ multipole vectors, but not close to the sum
vector of multipoles up to $\ell=8$. They are close to the ecliptic pole and are
nearly orthogonal to the CMB dipole apexes; however, they drift upon the increase
of the temperature threshold interval. The effect weakens, but does not disappear
completely when the $\ell =2$ or $\ell =2,3$ modes are extracted.
There are visible mirroring properties, even if not perfect.

The association to these inhomogeneities with mirroring features either to the
ecliptic, or to the dipole or to the l=3 multipole, i.e. to unknown interplay of
interplanetary and Galactic foregrounds is certainly a central issue. In that case,
however, one is led to reconsider such non-cosmological contributions in the low
multipoles. For now, this seems one of the strong alternatives.

What if the described features are nevertheless cosmological? Do
there exist fundamental mechanisms producing such an {\em
approximate} mirror symmetry of CMB temperature fluctuations with
respect to some fixed plane, apart from a pure chance? The answer is
positive, and the simplest possibility is presented by the
non-trivial spatial topology of the Universe of the $T^1$ type, i.e.
with one of spatial coordinates, say $z$, being identified: $z\equiv
z+L$. This topology may be considered as a limiting case of the
Universe with compact flat spatial sections having the $T^3$
topology if the identification scales $L_1,L_2$ along two other
spatial coordinates are much more than $L$ (the slab topology), see
\cite{ZS84,GK89,L04} for the discussions of quantum creation of the
early universe with such topology. Another possibility is to have
compact positively curved spatial sections with the radius of
curvature much exceeding the present cosmological horizon $R_{hor}$
(so that $\Omega_{tot}-1\ll 1$) and with the topology $S^3/Z_N,~N\gg 1$.

As was shown in \cite{S93}, in the case of the $T^1$ spatial
topology oriented along the $z$ axis, a large-angle pattern of a CMB
temperature anisotropy $\Delta T(\theta,\phi)$ is a sum of two
terms. The first of them has the exact plane (mirror) symmetry with
respect to the $(x,y)$ plane, namely
\begin{equation}\label{mirror}
\Delta T(\theta,\phi)= \Delta T (\pi-\theta,\phi).
\end{equation}
It originates from the Sachs-Wolfe effect at the large scattering
surface from density perturbations which does not depend on $z$. The
second term represents the remaining part of anisotropy and does not
have any symmetry at all. However, for $a_0L$ of the order of
$R_{\rm hor}$ or slightly more, where $a_0=a(t_0)$ is the present
scale factor of a Friedmann-Robertson-Walker cosmological model, the
latter term should be somehow suppressed since the Sachs-Wolfe
contribution to it from the last scattering surface comes from
perturbations having wave vectors with $|{\bf k}|\ge 2\pi/L$. That
is why we can expect a total large-angle pattern of $\Delta T$ to
have an {\em approximate} mirror symmetry in this case which should
quickly disappear at smaller angles.\footnote{An additional effect
worsening this symmetry at large angles is a contribution from the
integrated Sachs-Wolfe effect at small redshifts due to a
cosmological constant.}

The first direct search for such an effect in \cite{OSS96} using the
COBE data with a negative result placed a lower bound of the
topological scale $a_0L>3h^{-1}~{\rm Gpc}~\approx 0.3R_{\rm hor}$ at
the $95\%$ confidence (the value of $R_{\rm hor}$ is given for
$\Omega_m=0.3,~\Omega_{\Lambda}=0.7$). Using the first-year WMAP
data and adding a different, "circles on the sky" method, this upper
bound was raised up to $\sim R_{hor}$ \cite{Oliv} and then even up
to $\approx 1.8 R_{hor}$ \cite{CSSK04}, see also
\cite{KAC06,CLMR06,ALST07}.

However, even $a_0L\ge 2R_{hor}$ (when the circles in the sky method
does not work at all) does not exclude observability of this effect,
if the $z$-independent large-scale part of 3D perturbations has a
sufficiently large amplitude. It should be emphasized that there is
{\em no} definite prediction of a relative amplitude of this 'zero'
mode of perturbations compared to a generic mode with $k_z\not= 0$
since there is no definite theory of how this topology arises.
Moreover, the recent re-analysis of this problem for the cubic $T^3$
topology ($L=L_1=L_2$) (for which there may be no approximate mirror
symmetry at all) using the three-year WMAP data \cite{AJLS07}
suggests that the lower limit in \cite{CSSK04} is too optimistic, so
even $a_0L=1.2 R_{\rm hor}$ is not excluded in this case (in
agreement with the lower limit obtained earlier in \cite{PK06}).
Clearly, a lower limit on $a_0L$ for the $T^1$ topology may not be
less than that for the more restrictive cubic $T^3$ topology, so
$a_0L\ge 1.2R_{\rm hor}$ seems to be still possible for the former
topology, too. Note also that a possible mirror symmetry of the
Universe was proposed in \cite{LM05}, however, with suppression of
even multipoles (odd point parity).

Thus, even though obtaining of secure constraints either on the torus topology
or the compactification scales needs further data and efforts, the large scale
non-perfect mirroring in the CMB maps revealed above, if cosmological, would
already indicate that the large-scale $z$-independent part of density perturbation
inside our cosmological horizon is anomalously large.

We thank Paolo de Bernardis for valuable discussions and help. AAS was
partially supported by the Research Program ``Astronomy" of the
Russian Academy of Sciences. He also thanks Prof. Remo Ruffini and
ICRANet, Pescara for hospitality during the final stage of this
project.

\end{document}